\documentclass[11pt,a4paper,english]{article}
\usepackage[T1]{fontenc}
\usepackage[latin9]{inputenc}
\usepackage{amsmath}
\usepackage{amssymb}
\usepackage{esint}

\makeatletter

\newcommand{\noun}[1]{\textsc{#1}}
\providecommand{\tabularnewline}{\\}

\usepackage{a4wide}
\makeatletter
\renewcommand\theequation{\hbox{\normalsize\arabic{section}.\arabic{equation}}}
\@addtoreset{equation}{section}
\renewcommand\thefigure{\hbox{\normalsize\arabic{section}.\arabic{figure}}}
\@addtoreset{figure}{section}
\renewcommand\thetable{\hbox{\normalsize\arabic{section}.\arabic{table}}}
\@addtoreset{table}{section}
\makeatother

\makeatother

\usepackage{babel}

\begin{document}

\title{Form factor perturbation theory from finite volume}

\author{G. Takács\\
HAS Theoretical Physics Research Group\\
H-1117 Budapest, Pázmány Péter sétány 1/A}

\date{29th September 2009}
\maketitle
\begin{abstract}
Using a regularization by putting the system in finite volume, we
develop a novel approach to form factor perturbation theory for non-integrable
models described as perturbations of integrable ones. This permits
to go beyond first order in form factor perturbation theory and in
principle works to any order. The procedure is carried out in detail
for double sine-Gordon theory, where the vacuum energy density and
breather mass correction is evaluated at second order. The results
agree with those obtained from the truncated conformal space approach.
The regularization procedure can also be used to compute other spectral
sums involving disconnected pieces of form factors such as those that
occur e.g. in finite temperature correlators.
\end{abstract}

\section{Introduction}

Form factor perturbation theory (FFPT) was developed in \cite{nonintegrable}
in order to evaluate quantities in a non-integrable model obtained
as a perturbation of an integrable one. Writing the Hamiltonian in
the form\[
H_{\mbox{nonintegrable}}=H_{\mbox{integrable}}+\lambda\int dx\Psi(t,x)\]
where $\Psi$ denotes the local perturbing field that breaks integrability,
the first order corrections to the vacuum (bulk) energy density and
particle masses are given as\begin{eqnarray*}
\delta\mathcal{E}_{vac} & = & \lambda\left\langle 0\right|\Psi\left|0\right\rangle _{\lambda=0}\\
\delta M_{ab}^{2} & = & 2\lambda F_{a\bar{b}}^{\Psi}\left(i\pi\,,\,0\right)\end{eqnarray*}
where \begin{equation}
F_{i_{1}\dots i_{n}}^{\Psi}\left(\vartheta_{1},\dots,\vartheta_{n}\right)=\langle0|\Psi(0,0)|A_{i_{1}}\left(\vartheta_{1}\right)\dots A_{i_{n}}\left(\vartheta_{n}\right)\rangle_{\lambda=0}\label{eq:psiff}\end{equation}
are the form factors of the perturbing operator calculated at the
integrable point $\lambda=0$ and $\bar{b}$ denotes the charge conjugate
of particle species $b$. It is possible to evaluate first order corrections
to the two-particle $S$ matrix and also the widths of decays induced
by the perturbation \cite{dgm}. 

The evaluation of higher order corrections has not been developed;
simple considerations along the lines of \cite{nonintegrable} lead
to divergent expressions. However there is no place for mass renormalization
by counter terms analogous to standard Feynman perturbation theory
because the operator (\ref{eq:psiff}) defined by the form factors
is already well-defined and physical. This is also confirmed by the
fact that when the non-integrable model is formulated using the truncated
conformal space approach pioneered by Yurov and Zamolodchikov \cite{yurov_zamolodchikov}
the mass gaps turn out to be finite and well-defined (the vacuum energy
can still have divergent contributions depending on the ultraviolet
weight of $\Psi$, but the differences between energy levels are all
finite).

This leads to the central idea of the paper: since the TCSA expression
for the relative energy levels in a finite volume $L$ is finite,
and the ingredients necessary to evaluate finite volume perturbation
theory can be determined from infinite volume form factors using the
approach developed in \cite{fftcsa1,fftcsa2}, one can write down
finite and well-defined analytic expressions for the perturbation
of finite volume energy levels (which are accurate up to so-called
residual finite size effects that decay exponentially with the volume
i.e. are non-analytic in $1/L$, i.e. valid to all orders in $1/L$).
Then the quantity of interest (bulk energy density or mass correction)
can be expressed directly in finite volume and the infinite volume
limit is taken only at the end of the calculation. This is the same
philosophy that was used to obtain the expression of finite temperature
one-point functions in \cite{fftcsa2}.

Eventually, since first order FFPT was used in \cite{fftcsa2} to
derive the expressions of one-particle and two-particle diagonal matrix
elements in finite volume, nothing new is to be gained from the application
of finite volume techniques at first order. However, we get new results
at second order: a consistent, generally valid way of calculating
corrections to vacuum energy density and particle masses. It can also
be extended to other quantities such as the S matrix, and to higher
order FFPT corrections as well.

It is best to consider a concrete model to develop and test this approach.
The model of choice is the double sine-Gordon model defined by the
Hamiltonian\begin{equation}
H_{\mathrm{DSG}}=\int dx\left(\frac{1}{2}\left(\partial_{t}\varphi\right)^{2}+\frac{1}{2}\left(\partial_{x}\varphi\right)^{2}-\mu:\cos\beta\varphi:+\lambda:\sin\frac{\beta}{2}\varphi:\right)\label{eq:dsgham}\end{equation}
understood as a perturbation of the massless free boson (which also
defines the normal ordering). It has attracted interest recently chiefly
because it is a prototype of non-integrable field theory which can
be understood by application of techniques developed in the context
of integrable field theories \cite{delfino_mussardo,dsg} and it also
has several interesting applications \cite{delfino_mussardo,nersesyan,bullough}
such as to the study of massive Schwinger model (two-dimensional quantum
electrodynamics) and a generalized Ashkin-Teller model (a quantum
spin system) which are discussed in \cite{delfino_mussardo}. 

The double sine-Gordon model (\ref{eq:dsgham}) can be considered
as a non-integrable perturbation of the integrable sine-Gordon field
theory obtained by setting $\lambda=0$ \cite{delfino_mussardo}.
Form factor perturbation theory was applied to the double sine-Gordon
model in \cite{dsg}; for the particular version in eqn. (\ref{eq:dsgham})
it was shown that the corrections to the breather masses vanish to
first order in $\lambda$. However later semiclassical considerations
\cite{semicl} seemed to contradict these naive expectations, yielding
mass corrections which were of first order in the coupling $\lambda$.
In \cite{dsg_mass} it was shown that a precise numerical determination
of the spectrum contradicts this conclusion and upholds the naive
picture obtained from form factor perturbation theory: i.e. there
are only second order corrections, and in fact all odd orders vanish
since the entire spectrum turns out to be even under $\lambda\rightarrow-\lambda$.
However, at that time the mass correction could not be calculated
theoretically due to the lack of FFPT beyond first order. Therefore
this model is an interesting testing ground for the present work,
and it is also made ideal by the absence of first order corrections
which makes comparison to numerical results easier. The numerical
results which are compared with the theoretical predictions are obtained
from TCSA which was first developed for the sine-Gordon model in \cite{frt1}
and generalized to the double sine-Gordon model in \cite{dsg}; to
achieve a better precision we use an improved version developed for
the work \cite{dsg_mass} and described therein.

\section{Bulk energy correction}

The general formula for second order corrections to energy levels
can be written as\[
\delta E_{i}=\sum_{k\neq i}\frac{\left|\left\langle i\left|H_{1}\right|k\right\rangle \right|^{2}}{E_{i}^{(0)}-E_{k}^{(0)}}\]
 where $E_{i}^{(0)}$ are the unperturbed energy eigenvalue corresponding
to the eigenstate $|i\rangle$ and $H_{1}$ is the perturbation to
the Hamiltonian. In our case \[
H_{1}=\lambda\int_{0}^{L}dx\,:\sin\frac{\beta}{2}\varphi:\]
where $L$ is the spatial volume of the system. With periodic boundary
conditions the matrix elements of $H_{1}$ vanish unless the momenta
of states $|i\rangle$ and $|k\rangle$ coincide; therefore when $i$
is taken to be the vacuum, only states with zero total momentum contribute.
In addition, the topological charge of $|k\rangle$ must also be zero,
otherwise the amplitude vanishes. Furthermore, $H_{1}$ is odd under
$C:\,\varphi\rightarrow-\varphi$ and so the $C$-parity of the contributing
state must be odd as well (the $n$th breather $B_{n}$ has $C$-parity
$(-1)^{n})$. Only contributions of breather states are necessary
to evaluate because our numerical data will come from a part of the
attractive regime of sine-Gordon theory where solitons are heavy and
contribute little to the summation over $k$, well below the available
numerical precision.

When the momentum of the state $|k\rangle$ is zero, \[
\left\langle 0\left|H_{1}\right|k\right\rangle =\lambda\left\langle 0\left|:\sin\frac{\beta}{2}\varphi(t,x):\right|k\right\rangle _{L}\]
is independent of $x$, therefore\begin{equation}
\delta E_{0}=-\lambda^{2}L^{2}\sum_{k\neq0}\frac{\left|\left\langle 0\left|:\exp i\frac{\beta}{2}\varphi(0,0):\right|k\right\rangle _{L}\right|^{2}}{E_{k}^{(0)}-E_{0}^{(0)}}\label{eq:2ndorderpt}\end{equation}
where the subscript $L$ designates finite volume matrix elements
(the two exponential terms in the sine give equal contributions due
to parity). For $\xi<1/3$%
\footnote{For the notations $M$ and $\xi$ cf. Appendix \ref{sec:Breather-form-factors}.%
}, the lowest lying states contributing to the sum are \begin{eqnarray*}
 &  & |B_{1}(0)\rangle\,,\quad|B_{3}(0)\rangle\,,\\
 &  & |B_{1}(\theta_{1})B_{2}(\theta_{2})\rangle\quad\mbox{with}\quad m_{1}\sinh\theta_{1}+m_{2}\sinh\theta_{2}=0\\
\mbox{and} &  & |B_{1}(\theta_{1})B_{1}(\theta_{2})B_{1}(\theta_{3})\rangle\quad\mbox{with}\quad m_{1}\sinh\theta_{1}+m_{1}\sinh\theta_{2}+m_{1}\sinh\theta_{3}=0\end{eqnarray*}
where \[
m_{k}=2M\sin\frac{\pi k\xi}{2}\]
are the breather masses, and the rapidities of the breathers are indicated
in parentheses. In details \begin{eqnarray*}
\delta E_{0}(L) & = & -\lambda^{2}L^{2}\frac{\left|\left\langle 0\left|:\exp i\frac{\beta}{2}\varphi(0,0):\right|B_{1}(0)\right\rangle _{L}\right|^{2}}{m_{1}}-\lambda^{2}L^{2}\frac{\left|\left\langle 0\left|:\exp i\frac{\beta}{2}\varphi(0,0):\right|B_{3}(0)\right\rangle _{L}\right|^{2}}{m_{3}}\\
 & - & \lambda^{2}L^{2}\sum_{\theta_{1}}\frac{\left|\left\langle 0\left|:\exp i\frac{\beta}{2}\varphi(0,0):\right|B_{1}(\theta_{1})B_{2}(\theta_{2})\right\rangle _{L}\right|^{2}}{(m_{1}\cosh\theta_{1}+m_{2}\cosh\theta_{2})}\\
 & - & \lambda^{2}L^{2}\sum_{\theta_{1},\theta_{2}}\frac{\left|\left\langle 0\left|:\exp i\frac{\beta}{2}\varphi(0,0):\right|B_{1}(\theta_{1})B_{1}(\theta_{2})B_{1}(\theta_{3})\right\rangle _{L}\right|^{2}}{(m_{1}\cosh\theta_{1}+m_{1}\cosh\theta_{2}+m_{1}\cosh\theta_{3})}+O\left(\mathrm{e}^{-\mu L}\right)+\dots\end{eqnarray*}
where the presence of correction terms decaying exponentially with
the volume is indicated, and the ellipsis denotes the terms corresponding
to further multi-particle states. The summations run over all distinct
solutions of the Bethe-Yang equations \[
Q_{k;a_{1}\dots a_{n}}(L|\theta_{1},\dots,\theta_{n})=m_{a_{k}}L\sinh\theta_{k}+\sum_{l\neq k}-i\log S_{a_{k}a_{l}}\left(\theta_{k}-\theta_{l}\right)=2\pi I_{k}\;,\qquad I_{k}\in\mathbb{Z}\]
that have total momentum zero. Using the results of \cite{fftcsa1}
one can write \begin{eqnarray*}
\left\langle 0\left|:\exp i\frac{\beta}{2}\varphi(0,0):\right|B_{1}(0)\right\rangle _{L} & = & \frac{F_{1}^{1/2}(0)}{\sqrt{\rho_{1}(L|0)}}+O\left(\mathrm{e}^{-\mu L}\right)\\
\left\langle 0\left|:\exp i\frac{\beta}{2}\varphi(0,0):\right|B_{3}(0)\right\rangle _{L} & = & \frac{F_{3}^{1/2}(0)}{\sqrt{\rho_{3}(L|0)}}+O\left(\mathrm{e}^{-\mu L}\right)\\
\left\langle 0\left|:\exp i\frac{\beta}{2}\varphi(0,0):\right|B_{1}(\theta_{1})B_{2}(\theta_{2})\right\rangle _{L} & = & \frac{F_{12}^{1/2}(\theta_{1},\theta_{2})}{\sqrt{\rho_{12}(L|\theta_{1},\theta_{2})}}+O\left(\mathrm{e}^{-\mu L}\right)\\
\left\langle 0\left|:\exp i\frac{\beta}{2}\varphi(0,0):\right|B_{1}(\theta_{1})B_{1}(\theta_{2})B_{1}(\theta_{3})\right\rangle _{L} & = & \frac{F_{111}^{1/2}(\theta_{1},\theta_{2})}{\sqrt{\rho_{111}(L|\theta_{1},\theta_{2},\theta_{3})}}+O\left(\mathrm{e}^{-\mu L}\right)\end{eqnarray*}
The density factors $\rho$ are obtained as\[
\rho_{i_{1}\dots i_{n}}(L|\theta_{1},\dots,\theta_{n})=\det\left\{ \frac{\partial Q_{k;a_{1}\dots a_{n}}}{\partial\theta_{l}}\right\} _{k,l=1,\dots,n}\]
In particular, for the one-particle densities we obtain\begin{equation}
\rho_{k}(L|\theta)=m_{k}L\cosh\theta\label{eq:oneptdensity}\end{equation}
The next step is to take the limit $L\rightarrow\infty$: the exponential
corrections can be dropped and the summations substituted with integrals\begin{eqnarray*}
\sum_{\theta_{1}} & \rightarrow & \int_{-\infty}^{\infty}\frac{d\theta_{1}}{2\pi}\tilde{\rho}_{12}(L|\theta_{1})\\
\sum_{\theta_{1},\theta_{2}} & \rightarrow & \int_{-\infty}^{\infty}\frac{d\theta_{1}}{2\pi}\frac{d\theta_{2}}{2\pi}\tilde{\rho}_{111}(L|\theta_{1},\theta_{2})\end{eqnarray*}
where $\tilde{\rho}$ denotes the density of zero-momentum states.
For the first integral, it can be obtained by inspecting the Bethe-Yang
equations\begin{eqnarray*}
m_{1}L\sinh\theta_{1}-i\log S_{12}(\theta_{1}-\theta_{2}) & = & 2\pi I_{1}\\
m_{2}L\sinh\theta_{2}-i\log S_{12}(\theta_{2}-\theta_{1}) & = & 2\pi I_{2}\end{eqnarray*}
with $S_{12}$ denoting the $B_{1}B_{2}$ $S$-matrix (cf. eqn. (\ref{eq:b1bksmat})).
The second equation is actually superfluous due to the zero-momentum
constraint $m_{1}\sinh\theta_{1}+m_{2}\sinh\theta_{2}=0$. Taking
the derivative of the first equation gives\begin{eqnarray*}
\tilde{\rho}_{12}(L|\theta_{1}) & = & m_{1}L\cosh\theta_{1}+\left(1+\frac{m_{2}\cosh\theta_{2}}{m_{1}\cosh\theta_{1}}\right)\Phi_{12}(\theta_{1}-\theta_{2})\\
 &  & \Phi_{12}(\theta)=-i\frac{\partial}{\partial\theta}\log S_{12}(\theta)\end{eqnarray*}
using the zero-momentum constraint during the differentiation. On
the other hand, the density factor $\rho_{12}$ is\begin{equation}
\rho_{12}(L|\theta_{1},\theta_{2})=m_{1}L\cosh\theta_{1}m_{2}L\cosh\theta_{2}+(m_{1}L\cosh\theta_{1}+m_{2}L\cosh\theta_{2})\Phi_{12}(\theta_{1}-\theta_{2})\label{eq:rho12}\end{equation}
and therefore\[
\frac{\tilde{\rho}_{12}(L|\theta_{1})}{\rho_{12}(L|\theta_{1},\theta_{2})}=\frac{1}{m_{2}L\cosh\theta_{2}}\]
A similar calculation yields\[
\frac{\tilde{\rho}_{111}(L|\theta_{1},\theta_{2})}{\rho_{111}(L|\theta_{1},\theta_{2},\theta_{3})}=\frac{1}{m_{1}L\cosh\theta_{3}}\]
The end result is that the correction is proportional to the volume
$L$, and therefore it represents a correction to the bulk energy
density\begin{eqnarray}
 &  & \delta\mathcal{E}=\frac{\delta E_{0}(L)}{L}=-\lambda^{2}\Bigg\{\frac{\left|F_{1}^{1/2}(0)\right|^{2}}{m_{1}^{2}}+\frac{\left|F_{3}^{1/2}(0)\right|^{2}}{m_{3}^{2}}\label{eq:bulkcontribs}\\
 &  & +\int_{-\infty}^{\infty}\frac{d\theta_{1}}{2\pi}\left.\left(\frac{\left|F_{12}^{1/2}(\theta_{1},\theta_{2})\right|^{2}}{(m_{1}\cosh\theta_{1}+m_{2}\cosh\theta_{2})m_{2}\cosh\theta_{2}}\right)\right|_{\theta_{2}=-\mathrm{arsinh}(m_{1}\sinh\theta_{1}/m_{2})}\nonumber \\
 &  & +\frac{1}{3!}\int_{-\infty}^{\infty}\int_{-\infty}^{\infty}\frac{d\theta_{1}}{2\pi}\frac{d\theta_{2}}{2\pi}\left.\left(\frac{\left|F_{111}^{1/2}(\theta_{1},\theta_{2},\theta_{3})\right|^{2}}{(m_{1}\cosh\theta_{1}+m_{1}\cosh\theta_{2}+m_{1}\cosh\theta_{3})m_{1}\cosh\theta_{3}}\right)\right|_{\theta_{3}=-\mathrm{arsinh}(\sinh\theta_{1}+\sinh\theta_{2})}\nonumber \\
 &  & +\dots\,\Bigg\}+O\left(\lambda^{4}\right)\nonumber \end{eqnarray}
where the form factor functions are defined in Appendix \ref{sec:Breather-form-factors}
(the combinatorial factor in the last integral takes into account
that states that only differ in the ordering of the rapidities are
eventually identical). 

The bulk energy density corrections can now be evaluated explicitly
in units of the soliton mass $M$. Introducing also the dimensionless
coupling \cite{dsg_mass} \[
t=\lambda M^{-2+\beta^{2}/16\pi^{2}}\]
we can write\[
\frac{\delta\mathcal{E}}{M^{2}}=-b_{2}t^{2}+O(t^{4})\]
The results of second order FFPT are summarized and compared to numerical
values extracted from TCSA in Table \ref{tab:Comparing-vacuum-energy}.
The accuracy of the data in the table corresponds to the estimated
precision of the TCSA results; at this level, the contribution of
the integral terms is negligible. The deviation between FFPT and TCSA
comes from two sources. For lower values of $R$, TCSA was observed
to converge slower, thereby limiting the accuracy of the numerical
determination. Albeit there exists a renormalization group method
for improving convergence \cite{tcsarg1,tcsarg2}, implementing it
comes with a cost (in terms of programming and running). It also does
not seem to gain much compared to the simple-minded approach of evaluating
bulk energy by the simpler method which was applied with success in
many previous examples \cite{dsg,dsg_mass}. Our method (also used
in \cite{dsg,dsg_mass}) is to find the scaling regime where the ground
state level is most linear (the region where its second derivative
in $L$ is smallest) and evaluate the slope of the line there. Similarly,
masses can be evaluated in the region where the gap between the appropriate
excited state and the ground state becomes closest to constant (found
by searching for the minimum of the first derivative) and taking the
value of the gap there as the approximate mass. 

For higher values of $R$, the spectrum of the theory becomes more
and more dense as the sine-Gordon model is increasingly attractive
(at the point $R=1.5$ there are three breather states in the spectrum,
while at $R=2.5$ there are already eleven of them), therefore there
are more multi-breather states to be included, and in addition there
are also states containing solitons.

\begin{table}
\begin{centering}
\begin{tabular}{|c|c|c|c|c|}
\hline 
$R=\sqrt{4\pi}/\beta$ & $1.5$ & $1.9$ & $2.2$ & $2.5$\tabularnewline
\hline
\hline 
$b_{2}$ (TCSA) & $0.81$ & $1.53$ & $2.21$ & $3.04$\tabularnewline
\hline 
$b_{2}$ (FFPT) & $0.82$ & $1.53$ & $2.20$ & $3.01$\tabularnewline
\hline 
$b_{s\bar{s}}$ (FFPT) & $0.0025$ & $4\times10^{-5}$ & $1\times10^{-6}$ & $1\times10^{-8}$\tabularnewline
\hline
\end{tabular}
\par\end{centering}

\caption{\label{tab:Comparing-vacuum-energy}Comparing vacuum energy density
from FFPT to TCSA numerics. The parameter $R$ is related to the compactification
radius of the ultraviolet limiting $c=1$ free boson conformal field
theory.}

\end{table}

To demonstrate that solitons contribute very little, let us also compute
the value of the first solitonic correction, which comes from the
soliton-antisoliton two-particle state. It can be written in a form
very similar to the $B_{1}B_{2}$ term: \[
-\lambda^{2}\int_{-\infty}^{\infty}\frac{d\theta}{2\pi}\frac{|F_{s\bar{s}}^{1/2}(\theta,-\theta)-F_{s\bar{s}}^{-1/2}(\theta,-\theta)|^{2}/4}{2M\cosh\theta\, M\cosh\theta}\]
where $F_{s\bar{s}}^{\pm1/2}$ is given in (\ref{eq:fsa}). The contribution
of this integral to $b_{2}$ is denoted $b_{s\bar{s}}$ and is shown
separately in table \ref{tab:Comparing-vacuum-energy}. The reason
for the smallness of this integral is that it has a very limited effective
support. The integrand is eventually symmetric in $\theta$, and form
factors generally vanish on threshold ($\theta=0$). On the other
hand, the form factor combination in the numerator exhibits an exponential
decay for large $\theta$\[
|F_{s\bar{s}}^{1/2}(\theta,-\theta)-F_{s\bar{s}}^{-1/2}(\theta,-\theta)|^{2}\sim\exp\left(-\frac{1-\xi}{\xi}\theta\right)\]
where $\xi<1$ in the attractive regime. Together with the denominator
this makes the integrand decay very fast with increasing $\theta$.
Similar behaviour happens in terms with larger number of particles,
ensuring the convergence of all multi-particle integrals involved.
Similar arguments hold also for the $B_{1}B_{2}$ term, but that is
made larger by the appearance of smaller masses ($m_{1},m_{2}$ instead
of $M$) in the denominator.

The issue of whether the summation over the states with increasing
number of particles implied by (\ref{eq:2ndorderpt}) converges is
more subtle since it is also necessary to take into account the various
numerical prefactors (form factor normalization etc.) entering the
individual contributions and is not considered here in detail. Just
as in the above calculation, the contributions can be naturally ordered
by the sum of the masses of the constituent particles in the intermediate
state, and explicit numerical evaluations support the observation
that they decrease very rapidly when going to more and more massive
states.

\section{Mass correction}

Let us now turn to evaluating the mass correction for the first breather
$B_{1}$. In finite volume, the $B_{1}$ one-particle state is just
the next energy level $|1\rangle$ above the vacuum $|0\rangle$in
the zero-momentum, zero topological charge sector. Therefore\begin{equation}
\delta E_{1}=-\lambda^{2}L^{2}\sum_{k\neq1}\frac{\left|\left\langle 1\left|:\exp i\frac{\beta}{2}\varphi(0,0):\right|k\right\rangle _{L}\right|^{2}}{E_{k}^{(0)}-E_{1}^{(0)}}\label{eq:level1}\end{equation}
and the correction to the mass gap is obtained by taking the difference
to the vacuum level:\[
\delta m_{1}=\lim_{L\rightarrow\infty}\delta E_{1}(L)-\delta E_{0}(L)\]

\subsection{Bulk contributions: a puzzle and its solution\label{sub:Bulk-contributions:-a}}

In particular, terms linear in the volume are expected to cancel,
leaving us with a finite correction to the mass gap. However, right
with the first term a serious problem appears. The first contribution
to (\ref{eq:level1}) is given by the vacuum state and can be written
as\begin{equation}
\lambda^{2}L^{2}\frac{\left|\left\langle B_{1}(0)\left|:\exp i\frac{\beta}{2}\varphi(0,0):\right|0\right\rangle _{L}\right|^{2}}{m_{1}\rho_{1}(L|0)}+O\left(\mathrm{e}^{-\mu L}\right)\mathop{\longrightarrow}_{L\rightarrow\infty}\lambda^{2}L\frac{\left|F_{1}^{1/2}(0)\right|^{2}}{m_{1}^{2}}\label{eq:wrongsign}\end{equation}
which has the wrong sign to cancel the corresponding contribution
to the vacuum energy density i.e. the first term in eqn. (\ref{eq:bulkcontribs}).

The puzzle can be solved by observing that the contribution from $B_{1}B_{1}$
two-particle states (which are naively of order $L^{0}$ for large
$L$) diverges as $L\rightarrow\infty$due to a disconnected piece.
Such divergent pieces are finite for $L<\infty$, and have a dependence
of $L$ to the power of the number of particles involved in the disconnected
part. In this case it leads to a piece proportional to $L$, and we
proceed to show that it gives the correct contributions to account
for the mismatch noted above. The corresponding term can be written
as \[
-\lambda^{2}L^{2}\sum_{\theta}\frac{\left|\left\langle B_{1}(0)\left|:\exp i\frac{\beta}{2}\varphi(0,0):\right|B_{1}(\theta)B_{1}(-\theta)\right\rangle _{L}\right|^{2}}{2m_{1}\cosh\theta-m_{1}}+O\left(\mathrm{e}^{-\mu L}\right)\]
and using the results of \cite{fftcsa1} this can be written as

\begin{equation}
-\lambda^{2}L^{2}\sum_{\theta}\frac{\left|F_{111}^{1/2}(i\pi,\theta,-\theta)\right|^{2}}{\rho_{1}(L|0)\rho_{11}(L|\theta,-\theta)(2m_{1}\cosh\theta-m_{1})}+O\left(\mathrm{e}^{-\mu L}\right)\label{eq:11contribfinvol}\end{equation}
with $\rho_{1}$ as in (\ref{eq:oneptdensity}) and\begin{eqnarray*}
\rho_{11}(L|\theta_{1},\theta_{2}) & = & m_{1}^{2}L^{2}\cosh\theta_{1}\cosh\theta_{2}+m_{1}L(\cosh\theta_{1}+\cosh\theta_{2})\Phi_{11}(\theta_{1}-\theta_{2})\\
 &  & \Phi_{11}(\theta)=-i\frac{\partial}{\partial\theta}\log S_{11}(\theta)\end{eqnarray*}
where \begin{equation}
S_{11}(\theta)=\frac{\sinh\theta+i\sin\pi\xi}{\sinh\theta-i\sin\pi\xi}\label{eq:b1b1ampl}\end{equation}
is the $B_{1}B_{1}$ scattering amplitude. A simple calculation similar
to that in the previous subsection gives the density of zero total
momentum states as \[
\tilde{\rho}_{11}(L|\theta)=m_{1}L\cosh\theta+2\Phi_{11}(2\theta)=\frac{\rho_{11}(L|\theta,-\theta)}{m_{1}L\cosh\theta}\]
Naive application of the infinite volume limit to (\ref{eq:11contribfinvol})
gives\[
-\lambda^{2}\int_{0}^{\infty}\frac{d\theta}{2\pi}\frac{\left|F_{111}^{1/2}(i\pi,\theta,-\theta)\right|^{2}}{m_{1}^{3}\cosh\theta(2\cosh\theta-1)}\]
However, the integral is divergent due to kinematical poles of the
form factor at $\theta=0$. The form factors in an integrable quantum
field theory satisfy a number of axioms (for the details we refer
to Smirnov's review \cite{Smirnov}), among which there is the kinematical
residue axiom of the form%
\footnote{This form of the axiom is valid for self-conjugate particles; for
charged particles it involves the charge conjugation matrix.%
} \begin{equation}
-i\mathop{\textrm{Res}}_{\theta=\theta^{'}}F_{n+2}^{\mathcal{O}}(\theta+i\pi,\theta^{'},\theta_{1},\dots,\theta_{n})_{i\, j\, i_{1}\dots i_{n}}=\left(1-\delta_{i\, j}\prod_{k=1}^{n}S_{i\, i_{k}}(\theta-\theta_{k})\right)F_{n}^{\mathcal{O}}(\theta_{1},\dots,\theta_{n})_{i_{1}\dots i_{n}}\label{eq:kinematicalaxiom}\end{equation}
which results in the following singularity\[
\left|F_{111}^{1/2}(i\pi,\theta,-\theta)\right|^{2}\sim\frac{16\left|F_{1}^{1/2}(0)\right|^{2}}{\theta^{2}}+O(\theta^{0})\]
using the fact that $S_{11}(0)=-1$ which expresses the Pauli exclusion
principle satisfied by the $B_{1}$ particles. Therefore one must
return to a more careful evaluation of the sum (\ref{eq:11contribfinvol}).
The quantization of $\theta$ in a finite volume is given by\begin{equation}
m_{1}L\sinh\theta+\delta_{11}(2\theta)=2\pi J\quad,\quad J\in\mathbb{N}+\frac{1}{2}\label{eq:by11}\end{equation}
where the quantum number is shifted by $-1/2$ due to the following
identification of the two-particle phase-shift:\[
S_{11}(\theta)=-\mathrm{e}^{i\delta_{11}(\theta)}\]
As a result we obtain that for fixed $J$ \[
\theta=\frac{2\pi J}{m_{1}L}\]
and the leading term in the sum (\ref{eq:11contribfinvol}) can be
written as\[
-\lambda^{2}L^{2}\sum_{J}\frac{16\left|F_{1}^{1/2}(0)\right|^{2}}{m_{1}L(m_{1}L)^{2}(2m_{1}-m_{1})}\left(\frac{m_{1}L}{2\pi J}\right)^{2}\]
Using the identity\[
\sum_{J\in\mathbb{N}+1/2}\frac{1}{J^{2}}=\frac{\pi^{2}}{2}\]
we obtain \[
-\lambda^{2}L\frac{2\left|F_{1}^{1/2}(0)\right|^{2}}{m_{1}^{2}}\]
which exactly compensates for the mismatch caused by the {}``wrong
sign'' in eqn. (\ref{eq:wrongsign}). The correction to the mass
term comes from the subleading $L^{0}$ term in the sum (\ref{eq:11contribfinvol}),
which requires a very careful evaluation that is carried out in Appendix
\ref{sec:Evaluation-of-disconnected}.

Moving to the next correction ($B_{3}$ term) to the bulk energy density
in (\ref{eq:bulkcontribs}) and keeping in mind the above example,
it is easy to see that its counterpart arises from the $B_{1}B_{3}$
contribution to (\ref{eq:level1}). Here we encounter a different
mechanism for the generation of the bulk term. The appropriate sum
to evaluate is \begin{eqnarray}
-\lambda^{2}L^{2}\sum_{\theta_{1}}\frac{\left|\left\langle B_{1}(0)\left|:\exp i\frac{\beta}{2}\varphi(0,0):\right|B_{1}(\theta_{1})B_{3}(\theta_{2})\right\rangle _{L}\right|^{2}}{(m_{1}\cosh\theta_{1}+m_{3}\cosh\theta_{2})-m_{1}}\label{eq:13sum}\\
\mbox{with }m_{1}\sinh\theta_{1}+m_{3}\sinh\theta_{2}=0\nonumber \end{eqnarray}
It turns out that due to $S_{13}(0)=+1$ the form factor\[
F_{113}^{1/2}(i\pi,\theta_{1},\theta_{2})\]
is regular as $\theta_{1}\rightarrow0$ (and therefore also $\theta_{2}\sim m_{1}\theta_{1}/m_{3}\rightarrow0$)
and so the above discrete sum converts directly to an integral of
the form\begin{equation}
-\lambda^{2}\int_{-\infty}^{\infty}\frac{d\theta_{1}}{2\pi}\left.\left(\frac{\left|F_{113}^{1/2}(i\pi,\theta_{1},\theta_{2})\right|^{2}}{(m_{1}\cosh\theta_{1}+m_{3}\cosh\theta_{2}-m_{1})m_{3}\cosh\theta_{2}}\right)\right|_{\theta_{2}=-\mathrm{arsinh}(m_{1}\sinh\theta_{1}/m_{3})}\label{eq:113integral}\end{equation}
However, the Bethe-Yang equations\begin{eqnarray*}
m_{1}L\sinh\theta_{1}-i\log S_{13}(\theta_{1}-\theta_{2}) & = & 2\pi I_{1}\\
m_{3}L\sinh\theta_{2}-i\log S_{13}(\theta_{2}-\theta_{1}) & = & 2\pi I_{2}\end{eqnarray*}
have the solution $\theta_{1}=\theta_{2}=0$ for $I_{1}=I_{2}=0$,
which is allowed due to $S_{13}(0)=+1$. Using the results of the
paper \cite{fftcsa2}, the finite volume matrix element can be written
as\begin{eqnarray*}
\left\langle B_{1}(0)\left|:\exp i\frac{\beta}{2}\varphi(0,0):\right|B_{1}(0)B_{3}(0)\right\rangle _{L} & = & \frac{1}{\sqrt{\rho_{13}(L|0,0)\rho_{1}(L|0)}}\left(F_{113}^{1/2}(i\pi,0,0)+m_{1}LF_{3}^{1/2}(0)\right)\end{eqnarray*}
with $\rho_{13}$ obtained from (\ref{eq:rho12}) by replacing the
index $2$ with $3$. The $\theta_{1}=0$ term of (\ref{eq:13sum})
therefore takes the form\begin{eqnarray}
 &  & -\lambda^{2}\left(\frac{\left|F_{3}^{1/2}(0)\right|^{2}}{m_{3}^{2}}L+2\Re\mathrm{e}\frac{F_{113}^{1/2}(i\pi,0,0)F_{3}^{1/2}(0)}{m_{1}m_{3}}-\frac{(m_{1}+m_{3})\left|F_{3}^{1/2}(0)\right|^{2}\Phi_{13}(0)}{m_{1}m_{3}^{2}}+O(L^{-1})\right)\nonumber \\
 &  & \Phi_{13}(\theta)=-i\frac{\partial}{\partial\theta}\log S_{13}(\theta)\label{eq:13p0}\end{eqnarray}
The first term is just the correct bulk contribution, the next two
terms must be added to the mass correction and the $L^{-1}$ corrections
can be discarded. 

The other bulk terms in (\ref{eq:bulkcontribs}), written as integrals,
are expected to follow from terms of (\ref{eq:level1}) with $B_{1}B_{1}B_{2}$
and $B_{1}B_{1}B_{1}B_{1}$ as intermediate state. However, their
evaluation is rather tedious and will not be pursued here. The corresponding
bulk parts in (\ref{eq:bulkcontribs}) are small and therefore it
is plausible that their contributions to the mass shift are small
as well, comparable to numerical accuracy of TCSA and the errors made
by neglecting other states. This assumption will be justified by the
later comparison to TCSA.

\subsection{Evaluating the mass correction}

Using the formulae in the previous subsection and the end result (\ref{eq:dm11})
of Appendix \ref{sec:Evaluation-of-disconnected}, the correction
to the first breather mass can be written as follows (the particle
composition of the contributing intermediate state is indicated below
each term):\begin{eqnarray}
\delta m_{1} & = & \underbrace{\delta m_{1}^{(11)}}_{B_{1}B_{1}}+\underbrace{\delta m_{1}^{(2)}}_{B_{2}}+\underbrace{\delta m_{1}^{(13)}}_{B_{1}B_{3}}+\underbrace{\delta m_{1}^{(22)}}_{B_{2}B_{2}}+\underbrace{\delta m_{1}^{(4)}}_{B_{4}}+\dots\label{eq:totmasscorr}\\
\delta m_{1}^{(11)} & = & -\lambda^{2}\int_{0}^{\infty}\frac{d\theta}{2\pi}\left(\frac{\left|F_{111}^{1/2}(i\pi,\theta,-\theta)\right|^{2}}{m_{1}^{3}\cosh\theta(2\cosh\theta-1)}-\frac{16\left|F_{1}^{1/2}(0)\right|^{2}}{m_{1}^{3}\sinh^{2}\theta\cosh\theta}\right)\nonumber \\
 & - & \lambda^{2}\times16\left|F_{1}^{1/2}(0)\right|^{2}\left(\frac{\Phi_{11}(0)}{4m_{1}^{3}}-\frac{1}{4m_{1}^{3}}\right)\nonumber \\
\delta m_{1}^{(2)} & = & -\lambda^{2}\frac{\left|F_{12}^{1/2}(i\pi,0)\right|^{2}}{m_{1}m_{2}(m_{2}-m_{1})}\nonumber \\
\delta m_{1}^{(4)} & = & -\lambda^{2}\frac{\left|F_{14}^{1/2}(i\pi,0)\right|^{2}}{m_{1}m_{4}(m_{4}-m_{1})}\nonumber \\
\delta m_{1}^{(13)} & = & -\lambda^{2}\left(2\Re\mathrm{e}\frac{F_{113}^{1/2}(i\pi,0,0)F_{3}^{1/2}(0)}{m_{1}m_{3}}-\frac{(m_{1}+m_{3})\left|F_{3}^{1/2}(0)\right|^{2}\Phi_{13}(0)}{m_{1}m_{3}^{2}}\right)\nonumber \\
 & - & \lambda^{2}\int_{-\infty}^{\infty}\frac{d\theta_{1}}{2\pi}\left.\left(\frac{\left|F_{113}^{1/2}(i\pi,\theta_{1},\theta_{2})\right|^{2}}{(m_{1}\cosh\theta_{1}+m_{3}\cosh\theta_{2}-m_{1})m_{3}\cosh\theta_{2}}\right)\right|_{\theta_{2}=-\mathrm{arsinh}(m_{1}\sinh\theta_{1}/m_{3})}\nonumber \\
\delta m_{1}^{(22)} & = & -\lambda^{2}\frac{1}{2!}\int_{-\infty}^{\infty}\frac{d\theta}{2\pi}\left(\frac{\left|F_{122}^{1/2}(i\pi,\theta,-\theta)\right|^{2}}{(2m_{2}\cosh\theta-m_{1})m_{2}\cosh\theta}\right)\nonumber \end{eqnarray}
(the evaluation of the terms $\delta m_{1}^{(2)}$, $\delta m_{1}^{(4)}$
and $\delta m_{1}^{(22)}$ proceeds by the already discussed methods;
there are no disconnected pieces in any of them). The mass correction
can be parametrized with the dimensionless coefficient $a_{2}$ defined
by\[
\frac{\delta m_{1}}{M}=-a_{2}t^{2}+O(t^{4})\]
which is compared to TCSA data in Table \ref{tab:Comparing-the-mass}.

\begin{table}
\begin{centering}
\begin{tabular}{|c|c|c|c|c|}
\hline 
$R=\sqrt{4\pi}/\beta$ & $1.6$ & $1.9$ & $2.2$ & $2.5$\tabularnewline
\hline
\hline 
$a_{2}$ (TCSA) & $3.8\pm0.3$ & $4.7\pm0.2$ & $6.1\pm0.1$ & $7.6\pm0.1$\tabularnewline
\hline 
$a_{2}$ (FFPT) & $3.66$ & $4.91$ & $6.23$ & $7.82$\tabularnewline
\hline
\end{tabular}
\par\end{centering}

\caption{\label{tab:Comparing-the-mass}Comparing the mass correction coefficient
$a_{2}$ from FFPT to TCSA numerics. The parameter $R$ is related
to the compactification radius of the ultraviolet limiting $c=1$
free boson conformal field theory. The values of $R$ are chosen to
lie in a range to ensure a sufficient precision for the TCSA determination,
for which an estimate of the numerical uncertainty is shown. FFPT
values from (\ref{eq:totmasscorr}) are reported with two decimal
places accuracy.}

\end{table}

\section{Conclusions}

In this paper it was shown how to use finite volume techniques to
go beyond first order in form factor perturbation theory. The second-order
corrections to vacuum energy and first breather mass was evaluated
in double sine-Gordon theory. In principle, the method works for higher
corrections, and for other quantities, such as the $S$ matrix%
\footnote{The evaluation of S matrix corrections can be carried out by calculating
the shifts of two-particle levels, which depend on the phase shift
in finite volume.%
} as well. 

The results of second order FFPT are in good agreement with numerical
data from TCSA. In addition, the regularization techniques developed
to evaluate disconnected contributions can also be used for the form
factor expansion of finite temperature two-point correlators. Eventually,
during the typing of this manuscript there appeared an independent
work by Essler and Konik \cite{esslerkonik}, which uses similar finite
volume techniques for correlators, and also introduces another, novel
infinite volume regularization procedure.

\subsection*{Acknowledgments}

This work was partially supported by the Hungarian OTKA grants K60040
and K75172. 

\appendix
\makeatletter \renewcommand{\theequation}{\hbox{\normalsize\Alph{section}.\arabic{equation}}} \@addtoreset{equation}{section} \renewcommand{\thefigure}{\hbox{\normalsize\Alph{section}.\arabic{figure}}} \@addtoreset{figure}{section} \renewcommand{\thetable}{\hbox{\normalsize\Alph{section}.\arabic{table}}} \@addtoreset{table}{section} \makeatother

\section{Breather form factors in sine-Gordon theory\label{sec:Breather-form-factors}}

To obtain matrix elements containing the first breather, one can analytically
continue the form factors of sinh-Gordon theory obtained in \cite{koubek_mussardo}
to imaginary values of the couplings. For the theory obtained by setting
$\lambda=0$ in (\ref{eq:dsgham}), the result reads\begin{eqnarray}
F_{\underbrace{{\scriptstyle 11\dots1}}_{n}}^{a}(\theta_{1},\dots,\theta_{n}) & = & \left\langle 0\left|\mathrm{e}^{ia\beta\varphi(0)}\right|B_{1}(\theta_{1})\dots B_{1}(\theta_{n})\right\rangle \nonumber \\
 & = & \mathcal{G}_{a}(\beta)\,[a]_{\xi}\,(i\bar{\lambda}(\xi))^{n}\,\prod_{i<j}\frac{f_{\xi}(\theta_{j}-\theta_{i})}{\mathrm{e}^{\theta_{i}}+\mathrm{e}^{\theta_{j}}}\, Q_{a}^{(n)}\left(\mathrm{e}^{\theta_{1}},\dots,\mathrm{e}^{\theta_{n}}\right)\label{eq:b1ffs}\end{eqnarray}
where $\xi=\beta^{2}/(8\pi-\beta^{2})$, \begin{eqnarray*}
Q_{a}^{(n)}(x_{1},\dots,x_{n}) & = & \det{[a+i-j]_{\xi}\,\sigma_{2i-j}^{(n)}(x_{1},\dots,x_{n})}_{i,j=1,\dots,n-1}\mbox{ if }n\geq2\\
Q_{a}^{(1)}=Q_{a}^{(2)} & = & 1\quad,\qquad[a]_{\xi}=\frac{\sin\pi\xi a}{\sin\pi\xi}\\
\bar{\lambda}(\xi) & = & 2\cos\frac{\pi\xi}{2}\sqrt{2\sin\frac{\pi\xi}{2}}\exp\left(-\int_{0}^{\pi\xi}\frac{dt}{2\pi}\frac{t}{\sin t}\right)\end{eqnarray*}
and \begin{eqnarray*}
f_{\xi}(\theta) & = & v(i\pi+\theta,-1)v(i\pi+\theta,-\xi)v(i\pi+\theta,1+\xi)v(-i\pi-\theta,-1)v(-i\pi-\theta,-\xi)v(-i\pi-\theta,1+\xi)\\
v(\theta,\zeta) & = & \prod_{k=1}^{N}\left(\frac{\theta+i\pi(2k+\zeta)}{\theta+i\pi(2k-\text{\ensuremath{\zeta}})}\right)^{k}\\
 &  & \times\exp\left\{ \int_{0}^{\infty}\frac{dt}{t}\left(-\frac{\zeta}{4\sinh\frac{t}{2}}-\frac{i\text{\ensuremath{\zeta}}\theta}{2\pi\cosh\frac{t}{2}}+\left(N+1-N\mbox{e}^{-2t}\right)\mbox{e}^{-2Nt+\frac{it\theta}{\pi}}\frac{\sinh\zeta t}{2\sinh^{2}t}\right)\right\} \end{eqnarray*}
gives the minimal $B_{1}B_{1}$ form factor%
\footnote{The formula for the function $v$ is in fact independent of $N$;
choosing $N$ large extends the width of the strip where the integral
converges and also speeds up convergence.%
}, while $\sigma_{k}^{(n)}$ denotes the elementary symmetric polynomial
of $n$ variables and order $k$ defined by\[
\prod_{i=1}^{n}(x+x_{i})=\sum_{k=0}^{n}x^{n-k}\sigma_{k}^{(n)}(x_{1},\dots,x_{n})\]
Furthermore\begin{eqnarray*}
\mathcal{G}_{a}(\beta)=\langle e^{ia\beta\varphi}\rangle & = & \left[\frac{M\sqrt{\pi}\Gamma\left(\frac{4\pi}{8\pi-\beta^{2}}\right)}{2\Gamma\left(\frac{\beta^{2}/2}{8\pi-\beta^{2}}\right)}\right]^{\frac{a^{2}\beta^{2}}{4\pi}}\\
 & \times & \exp\left\{ \int_{0}^{\infty}\frac{dt}{t}\left[\frac{\sinh^{2}\left(\frac{a}{4\pi}t\right)}{2\sinh\left(\frac{\beta^{2}}{8\pi}t\right)\cosh\left(\left(1-\frac{\beta^{2}}{8\pi}\right)t\right)\sinh t}-\frac{a^{2}\beta^{2}}{4\pi}e^{-2t}\right]\right\} \end{eqnarray*}
is the exact vacuum expectation value of the exponential field \cite{exact_vevs},
with $M$ denoting the soliton mass related to the coupling $\mu$
via \cite{mass_scale} \begin{equation}
\mu=\frac{2\Gamma(\Delta)}{\pi\Gamma(1-\Delta)}\left(\frac{\sqrt{\pi}\Gamma\left(\frac{1}{2-2\Delta}\right)M}{2\Gamma\left(\frac{\Delta}{2-2\Delta}\right)}\right)^{2-2\Delta}\qquad,\qquad\Delta=\frac{\beta^{2}}{8\pi}\label{eq:mass_scale}\end{equation}
Formula (\ref{eq:b1ffs}) also coincides with the result given in
\cite{lukyformfactors}.

Form factors containing higher breathers can be obtained using that
$B_{n}$ is a bound state of $B_{1}$ and $B_{n-1}$; therefore sequentially
fusing $n$ adjacent first breathers gives $B_{n}$. Following the
lines of reasoning of Appendix A of the paper \cite{resonances} one
obtains \begin{eqnarray}
 &  & F_{k_{1}\dots k_{r}nl_{1}\dots l_{s}}^{a}(\theta_{1},\dots,\theta_{r},\theta,\theta_{1}',\dots,\theta_{s}')=\label{eq:higherbff}\\
 &  & \left\langle 0\left|\mathrm{e}^{ia\beta\Phi(0)}\right|B_{k_{1}}(\theta_{1})\dots B_{k_{r}}(\theta_{r})B_{n}(\theta)B_{l_{1}}(\theta_{1}')\dots B_{l_{s}}(\theta_{s}')\right\rangle =\gamma_{11}^{2}\gamma_{12}^{3}\dots\gamma_{1n-1}^{n}\nonumber \\
 &  & \times F_{k_{1}\dots k_{r}\underbrace{{\scriptstyle 11\dots1}}_{n}l_{1}\dots l_{s}}^{a}\left(\theta_{1},\dots,\theta_{r},\theta+\frac{1-n}{2}i\pi\xi,\theta+\frac{3-n}{2}i\pi\xi,\dots,\theta+\frac{n-1}{2}i\pi\xi,\theta_{1}',\dots,\theta_{s}'\right)\nonumber \end{eqnarray}
where\[
\gamma_{1k}^{k+1}=\sqrt{\frac{2\tan\frac{k\pi\xi}{2}\tan\frac{(k+1)\pi\xi}{2}}{\tan\frac{\pi\xi}{2}}}\]
is the $B_{1}B_{k}\rightarrow B_{k+1}$ coupling, defined as the residue
of the appropriate scattering amplitude:\begin{eqnarray}
i\left(\gamma_{1k}^{k+1}\right)^{2} & = & \mathop{\mbox{Res}}_{\theta=\frac{i\pi(k+1)\xi}{2}}S_{1k}(\theta)\nonumber \\
S_{1k}(\theta) & = & \frac{\sinh\theta+i\sin\frac{\pi(k+1)\xi}{2}}{\sinh\theta-i\sin\frac{\pi(k+1)\xi}{2}}\frac{\sinh\theta+i\sin\frac{\pi(k-1)\xi}{2}}{\sinh\theta-i\sin\frac{\pi(k-1)\xi}{2}}\label{eq:b1bksmat}\end{eqnarray}
Using the results of \cite{lukyformfactors} we also quote here the
simplest solitonic form factor, which is needed in the main text\begin{equation}
F_{s\bar{s}}^{\pm1/2}(\theta_{1},\theta_{2})=\mathcal{G}_{a}(\beta)\frac{G(\theta_{2}-\theta_{1})}{G(-i\pi)}\frac{2i\mathrm{e}^{\mp\frac{\theta+i\pi}{2\xi}}}{\xi\sinh\left(\frac{\theta+i\pi}{\xi}\right)}\label{eq:fsa}\end{equation}
where \begin{eqnarray*}
G(\theta) & = & i\mathcal{C}_{1}\sinh\frac{\theta}{2}\exp\left(\int_{0}^{\infty}\frac{dt}{t}\frac{\sinh^{2}t(1-\frac{i\theta}{\pi})\sinh t(\xi-1)}{\sinh2t\,\cosh t\,\sinh t\xi}\right)\\
\mathcal{C}_{1} & = & \exp\left(-\int_{0}^{\infty}\frac{dt}{t}\frac{\sinh^{2}\frac{t}{2}\sinh t(\xi-1)}{\sinh2t\,\cosh t\,\sinh t\xi}\right)\end{eqnarray*}

\section{Evaluation of disconnected contributions\label{sec:Evaluation-of-disconnected}}

To obtain $\delta m_{1}^{(11)}$ we need to evaluate the $O(L^{0})$
of the sum (\ref{eq:11contribfinvol}), which is

\[
-\lambda^{2}L^{2}\sum_{\theta}\frac{\left|F_{111}^{1/2}(i\pi,\theta,-\theta)\right|^{2}}{\rho_{1}(L|0)\rho_{11}(L|\theta,-\theta)(2m_{1}\cosh\theta-m_{1})}+O\left(\mathrm{e}^{-\mu L}\right)\]
Using\[
\left|F_{111}^{1/2}(i\pi,\theta,-\theta)\right|^{2}\sim\frac{16\left|F_{1}^{1/2}(0)\right|^{2}}{\theta^{2}}+O(\theta^{0})\]
we can subtract the singular piece to write\[
-\lambda^{2}L^{2}\sum_{\theta}\left(\frac{\left|F_{111}^{1/2}(i\pi,\theta,-\theta)\right|^{2}}{\rho_{1}(L|0)\rho_{11}(L|\theta,-\theta)(2m_{1}\cosh\theta-m_{1})}-\frac{16\left|F_{1}^{1/2}(0)\right|^{2}}{\sinh^{2}\theta\rho_{1}(L|0)\rho_{11}(L|\theta,-\theta)m_{1}}\right)\]
which can be readily converted for $L\rightarrow\infty$ to the following
integral:\[
-\lambda^{2}\int_{0}^{\infty}\frac{d\theta}{2\pi}\left(\frac{\left|F_{111}^{1/2}(i\pi,\theta,-\theta)\right|^{2}}{m_{1}^{3}\cosh\theta(2\cosh\theta-1)}-\frac{16\left|F_{1}^{1/2}(0)\right|^{2}}{m_{1}^{3}\sinh^{2}\theta\cosh\theta}\right)\]
Therefore what we need is the $O(L^{0})$ part of the subtracted term
\begin{equation}
\lambda^{2}L^{2}\sum_{\theta}\frac{16\left|F_{1}^{1/2}(0)\right|^{2}}{\sinh^{2}\theta\rho_{1}(L|0)\rho_{11}(L|\theta,-\theta)m_{1}}\label{eq:subtraction}\end{equation}
where\[
\rho_{1}(L|0)=m_{1}L\:,\quad\rho_{11}(L|\theta,-\theta)=m_{1}^{2}L^{2}\cosh^{2}\theta+2m_{1}L\cosh\theta\Phi_{11}(2\theta)\]
According to eqn. (\ref{eq:by11}) the rapidity is quantized as\[
m_{1}L\sinh\theta+\delta_{11}(2\theta)=2\pi J\]
where $J$ is a positive half-integer. Using the arguments of Subsection
\ref{sub:Bulk-contributions:-a} \[
\lambda^{2}L^{2}\sum_{J}\frac{16\left|F_{1}^{1/2}(0)\right|^{2}}{m_{1}^{2}L(m_{1}L)^{2}}\left(\frac{m_{1}L}{2\pi J}\right)^{2}=\lambda^{2}L^{2}\sum_{\theta}\frac{16\left|F_{1}^{1/2}(0)\right|^{2}}{m_{1}^{2}L(m_{1}L)^{2}}\left(\frac{1}{\sinh\theta+\frac{\delta_{11}(2\theta)}{m_{1}L}}\right)^{2}=\lambda^{2}L\,\frac{2\left|F_{1}^{1/2}(0)\right|^{2}}{m_{1}^{2}}\]
is just the $O(L)$ part of the sum (\ref{eq:subtraction}), which
can be explicitly subtracted without affecting the $O(L^{0})$ part.
Thus the $O(L^{0})$ term of (\ref{eq:subtraction}) can be obtained
as the $L\rightarrow\infty$ limit of\begin{eqnarray*}
 &  & 16\frac{\lambda^{2}\left|F_{1}^{1/2}(0)\right|^{2}}{m_{1}^{3}}\sum_{\theta}\left(\frac{1}{\sinh^{2}\theta\cosh\theta\left(m_{1}L\cosh\theta+2\Phi_{11}(2\theta)\right)}-\frac{1}{m_{1}L\sinh^{2}\theta}\right)\\
 & + & 16\frac{\lambda^{2}\left|F_{1}^{1/2}(0)\right|^{2}}{m_{1}^{4}L}\sum_{\theta}\left(\frac{1}{\sinh^{2}\theta}-\frac{1}{\left(\sinh\theta+\frac{\delta_{11}(2\theta)}{m_{1}L}\right)^{2}}\right)\end{eqnarray*}
where an intermediate subtraction was inserted to simplify the evaluation.
The second sum can be written\[
\sum_{\theta}\left(\frac{1}{\sinh^{2}\theta}-\frac{1}{\left(\sinh\theta+\frac{\delta_{11}(2\theta)}{m_{1}L}\right)^{2}}\right)=\sum_{J}\frac{m_{1}^{2}L^{2}}{4\pi^{2}J^{2}}\left(\frac{2\delta_{11}(2\theta)}{m_{1}L\sinh\theta}+\left(\frac{\delta_{11}(2\theta)}{m_{1}L\sinh\theta}\right)^{2}\right)\]
Now the singular part is in the $J^{-2}$ prefactor, and the remainder
is finite for any fixed $J$ when $L\rightarrow\infty$. Since in
the infinite volume limit $\theta\rightarrow0$ for any fixed $J$,
and the summation over $J$ is uniformly convergent, permitting to
exchange the limit with the sum, we can write\[
\sum_{\theta}\left(\frac{1}{\sinh^{2}\theta}-\frac{1}{\left(\sinh\theta+\frac{\delta_{11}(2\theta)}{m_{1}L}\right)^{2}}\right)=\sum_{J}\frac{m_{1}L}{4\pi^{2}J^{2}}4\Phi_{11}(0)+O(L^{0})=\frac{m_{1}L\Phi_{11}(0)}{2}+O(L^{0})\]
For the first sum we obtain\begin{eqnarray*}
 &  & \sum_{\theta}\left(\frac{1}{\sinh^{2}\theta\cosh\theta\left(m_{1}L\cosh\theta+2\Phi_{11}(2\theta)\right)}-\frac{1}{m_{1}L\sinh^{2}\theta}\right)=\\
 &  & -\sum_{\theta}\frac{1}{\sinh^{2}\theta}\left(\frac{1-1/\cosh^{2}\theta}{m_{1}L}+2\frac{\Phi_{11}(2\theta)}{m_{1}^{2}L^{2}\cosh^{3}\theta}+O(L^{-3})\right)=\\
 &  & -\frac{1}{4}-\frac{\Phi_{11}(0)}{4}+O(L^{-1})\end{eqnarray*}
where in the first term we used\[
\sum_{\theta}\frac{1-1/\cosh^{2}\theta}{m_{1}L\sinh^{2}\theta}=\int_{0}^{\infty}\frac{d\theta}{2\pi}(m_{1}L\cosh\theta+\Phi_{11}(2\theta))\frac{1}{m_{1}L\cosh^{2}\theta}=\frac{1}{4}+O(L^{-1})\]
(the integrand is non-singular at $\theta=0$, hence it can be evaluated
by a density integral), while for the second term: \[
\sum_{\theta}\frac{1}{\sinh^{2}\theta}\frac{2\Phi_{11}(2\theta)}{m_{1}^{2}L^{2}\cosh^{3}\theta}=\sum_{\theta}\frac{2\Phi_{11}(0)}{m_{1}^{2}L^{2}\sinh^{2}\theta}+O(L^{-1})=\sum_{J}\frac{2\Phi_{11}(0)}{4\pi^{2}J^{2}}+O(L^{-1})=\frac{\Phi_{11}(0)}{4}+O(L^{-1})\]
where again, all parts non-singular at $\theta=0$ were evaluated
at the origin. Collecting all the pieces we obtain that the subtracted
part (\ref{eq:subtraction}) equals\[
\lambda^{2}L^{2}\sum_{\theta}\frac{16\left|F_{1}^{1/2}(0)\right|^{2}}{\sinh^{2}\theta\rho_{1}(L|0)\rho_{11}(L|\theta,-\theta)m_{1}}=\lambda^{2}L\,\frac{2\left|F_{1}^{1/2}(0)\right|^{2}}{m_{1}^{2}}+\lambda^{2}\times16\left|F_{1}^{1/2}(0)\right|^{2}\left(\frac{\Phi_{11}(0)}{4m_{1}^{3}}-\frac{1}{4m_{1}^{3}}\right)+O(L^{-1})\]
and therefore\begin{eqnarray}
\delta m_{1}^{(11)} & = & -\lambda^{2}\int_{0}^{\infty}\frac{d\theta}{2\pi}\left(\frac{\left|F_{111}^{1/2}(i\pi,\theta,-\theta)\right|^{2}}{m_{1}^{3}\cosh\theta(2\cosh\theta-1)}-\frac{16\left|F_{1}^{1/2}(0)\right|^{2}}{m_{1}^{3}\sinh^{2}\theta\cosh\theta}\right)\label{eq:dm11}\\
 &  & -\lambda^{2}\times16\left|F_{1}^{1/2}(0)\right|^{2}\left(\frac{\Phi_{11}(0)}{4m_{1}^{3}}-\frac{1}{4m_{1}^{3}}\right)\nonumber \end{eqnarray}

\end{document}